\documentclass{article}
\usepackage{geometry}
\usepackage{multicol}
\usepackage{abstract}
\usepackage{authblk} 
\usepackage{amsmath}                 
\usepackage{amsfonts}
\usepackage{hyperref}
\usepackage{ragged2e}
\usepackage{graphicx}                

\usepackage{booktabs}                
\usepackage{longtable} 
\usepackage{tabularx}

\usepackage{xcolor}
\usepackage{adjustbox} 
\usepackage{float}  
\usepackage{caption}
\usepackage{longtable}
\title{\fontsize{23pt}{3pt}\selectfont{Image Processing and Analysis of Multiple Wavelength Astronomical Data Using Python Tools}}

\author[1]{\fontsize{12pt}{14pt}\selectfont \textit{Tanmoy Bhowmik}}
\author[2]{\fontsize{12pt}{14pt}\selectfont \textit{MD Fardin Islam}}
\author[3]{\fontsize{12pt}{14pt}\selectfont \textit{Kazi Nusrat Tasneem}}
\author[4]{\fontsize{12pt}{14pt}\selectfont \textit{Rantideb Roy}}
\author[5]{\fontsize{12pt}{14pt}\selectfont \textit{Rownok Shahariar}}

\affil[1]{\small Dept. Of Physics, Shahjalal University of Science and Technology, Sylhet }
\affil[2]{\small Dept. Of Physical Sciences, Independent University, Bangladesh }
\affil[3]{\small Dept. Of Computer Science and Engineering, Shahjalal University of Science and Technology, Sylhet }
\affil[4]{\small Dept. Of Computer Science and Engineering, Shahjalal University of Science and Technology, Sylhet }
\affil[5]{\small Dept. Of Physics, Govt. Edward College, Pabna }
\date{}

\begin{document}

\maketitle
\begin{abstract}
\fontsize{11pt}{4}\selectfont
\noindent
 We developed a Python based framework for astronomical image processing and analysis. Astronomical image loading, normalizing, stacking, and filtering processes represent visible range images from grayscale. Besides, the blending process helps to analyze the image of multiple wavelengths in the visible range. The methods take advantage of include median filtering for noise reduction, unsharp masking for sharpening details, and intensity normalization techniques. The detailed analysis of pixel intensity distributions and applying Gaussian fitting  to variations across different wavelength bands. These methods highlight Python as a valuable tool for astronomers.
\end{abstract}

\begin{flushleft}
\hspace{0.97cm}\footnotesize \textbf{Keywords}: Normalization, FITS, Noise Reduction, Composite, Blending, Median Filter, Gaussian, Stacking, RGB, CCD.
\end{flushleft}
    
\begin{multicols}{2}
\begin{justify}
\section{Introduction}
The development of astronomical image processing has come a long way. Astronomical observations have changed because of the use of charge-coupled devices (CCDs) in telescopes. Observations made in astronomy have shifted since telescopes started embracing charge-coupled devices that enable astronomers to photograph light accurately from  objects. This device (Fig~\ref{Fig 1}) converts incoming photons into electronic signals, creating a digital image of the observed scene.
\begin{figure}[H]
    \centering
    \includegraphics[width=9.2cm,height=6cm]{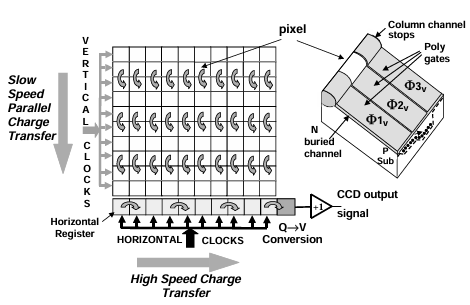}
    \caption{CCD Architecture (\cite{ref9})} 
    \label{Fig 1}
\end{figure}
\begin{figure}[H]
    \centering
    \includegraphics[width=7.5cm,height=6cm]{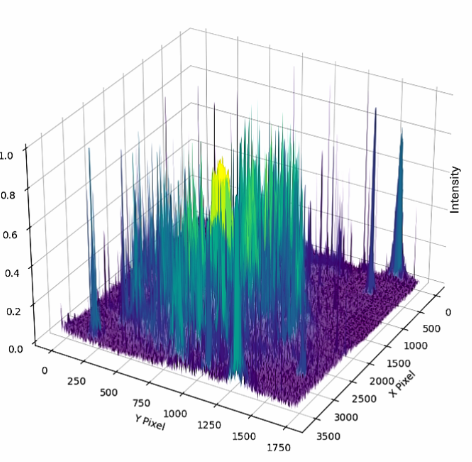}
    \caption{Noisy Image}
    \label{Fig2}
\end{figure}
 This process has many steps, from the initial capture of photons to the final production of a Flexible Image Transport System (FITS) file, the standard format for storing and sharing astronomical data. When light from a distant star or galaxy hits the CCD, it generates electron charges proportional to the intensity of the light. Each pixel on the CCD represents a specific location in the sky. The total collected charges are read and digitized to form a raw image. This raw data is enhanced by first creating a stack of multiple exposures to enhance signal strength over noise (Fig~\ref{Fig2}) levels. Applying the square root function provides a good balance between compressing the high-intensity values (bright areas) and expanding the low intensity values (dim areas). This is especially important in astronomy, where the dynamic range is vast for some stars may be extremely bright while other structures such as nebulae, distant galaxies  are faint. The square root helps to make dim objects more visible without completely washing out bright objects (\cite{ref15}).
\newline
\newline
 Interference is always a problem in astronomical imaging. Variations in CCD images that result from noises are due to thermal noise, cosmic rays, and readout noise (Fig~\ref{Fig2}). One of the key components in analyzing true signals involves the reduction of distorted noise signals and therefore, efficient methods in noise reduction is very important. For instance, median filtering is used to remove “salt and pepper” noise commonly caused by individual bright or dark pixel (\cite{ref3}). This method provide that though the noise is minimized, the outlines and features of objects in space are still well defined. 
\newline
\newline
 In addition to the Gaussian filtering in this process as in unsharp masking where in it helps generate a “mask” of low frequency detailing. If this mask is then taken away from the original image and the high frequency is boosted, what is obtained is a much sharper picture. Altogether, the mentioned methods improve the quality and contrast of astronomical images. 
 \newline
 \newline
 The significant of image processing is the formation of the  combined images, especially if dealing with multi-wavelength data. Combining two or more images in different wavelength bands like optical, infrared and X-ray give a higher resolution and more detailed picture of the phenomena. Such a technique allows astronomers to sort out details that can go unnoticed at a single wavelength, allowing a deep probe into the physical processes driving the objects.
 \newline
 \newline
 The discovery of a powerful tool in the form of Python has extended the frontiers of data analysis and scientific computing mainly in astronomy and image processing.  The various libraries; NumPy, Matplotlib, Astropy and Scikit-Image of Python enable easy handling of large data sets in image processing and visualization. Compared to other specialized astronomical software, Python offers flexibility and integration with various scientific computing tools, making it an essential tool for modern astronomical research. 
\end{justify}
\section{Methodology}
\label{sec:method}
{\small 
\subsection{Data Loading }
In the image processing and analysis, we used images in the FITS (Flexible Image Transport System) format. The procedure for loading image data involves opening the corresponding FITS files for each color channel: green, red, and blue. This process can be represented as follows:
\begin{equation*}
G = \text{fits.open}(\text{path\_to\_green\_image.fits}) 
\end{equation*}
\begin{equation*}
R = \text{fits.open}(\text{path\_to\_red\_image.fits}) 
\end{equation*}
\begin{equation*}
B = \text{fits.open}(\text{path\_to\_blue\_image.fits}) 
\end{equation*}

\subsection{Inside the R-G-B Matrices}

Each of the \textbf{R}, \textbf{G}, and \textbf{B} files in an image represents a 2D matrix of pixel intensity values for the Red, Green, and Blue channels, respectively. These matrices can be represented as:

\begin{equation*}
R = \left[ 
    \begin{array}{cccc}
    R_{11} & R_{12} & \cdots & R_{1n} \\
    R_{21} & R_{22} & \cdots & R_{2n} \\
    \vdots & \vdots & \ddots & \vdots \\
    R_{m1} & R_{m2} & \cdots & R_{mn}
    \end{array} 
\right]
\end{equation*}

\begin{equation*}
G = \left[
    \begin{array}{cccc}
    G_{11} & G_{12} & \cdots & G_{1n} \\
    G_{21} & G_{22} & \cdots & G_{2n} \\
    \vdots & \vdots & \ddots & \vdots \\
    G_{m1} & G_{m2} & \cdots & G_{mn}
    \end{array} 
\right]
\end{equation*}

\begin{equation*}
B = \left[ 
    \begin{array}{cccc}
    B_{11} & B_{12} & \cdots & B_{1n} \\
    B_{21} & B_{22} & \cdots & B_{2n} \\
    \vdots & \vdots & \ddots & \vdots \\
    B_{m1} & B_{m2} & \cdots & B_{mn}
    \end{array} 
\right]
\end{equation*}
} 
Where $R_{ij}$, $G_{ij}$, $B_{ij}$  
  represent the intensity of the red, green, and blue channels at pixel $(i,j)$. and $m \times n$ is the resolution of the image (i.e., the number of rows $m$ and columns $n$ of pixels). Each value $R_{ij}$, $G_{ij}$, $B_{ij}$ represents the intensity of the respective color channel at a specific pixel.
  
\subsection{Data Extraction}
 The process involves extracting the image data arrays by the primary header of the FITS file, which contains the parameters corresponding to the primary Header Data Unit (HDU). The extracted data arrays for the respective color channels can be represented as :
\begin{equation*}
I_R = R[0].\text{data} 
\end{equation*}
\begin{equation*}
I_G = G[0].\text{data} 
\end{equation*}
\begin{equation*}
I_B = B[0].\text{data} 
\end{equation*}
\subsection{Square Root Normalization of Red, Green, and Blue Images}

Square root normalization refers to applying the square root function to pixel values before displaying them. If a pixel value is denoted as $I$, then the normalized value $I_{\text{norm}}$ is computed as:
\[
I_{\text{norm}} = \sqrt{I}
\]
This transformation is nonlinear and tends to compress the higher range of values while spreading out lower values. 
\newline
\newline
Mathematically, the square root function grows slower than the linear function, especially for larger values of $I$. As we can see, for larger values of $I$, the difference between the transformed values becomes smaller. This leads to a more balanced representation of both faint and bright regions in the image.

\subsection{Stacking the R-G-B Matrices}
\begin{justify}
Stacking the \textbf{normalized} \textbf{R}, \textbf{G}, and \textbf{B} images involve combining them into a single 3D matrix (or tensor) where each element of this tensor contains three values, representing the red, green, and blue intensity for each pixel. 
\newline
\newline
If we take the individual matrices $R$, $G$, $B$ of size $m \times n$, stacking (Fig~\ref{fig3}) them results in a 3D tensor $\text{RGB}$ of size $m \times n \times 3$.

Where the third dimension holds the three color channels for each pixel:

\[
\fontsize{5}{1}
\text{RGB} = \left[ 
    \begin{array}{cccc}
    (R_{11}, G_{11}, B_{11}) & (R_{12}, G_{12}, B_{12}) & \cdots & (R_{1n}, G_{1n}, B_{1n}) \\
    (R_{21}, G_{21}, B_{21}) & (R_{22}, G_{22}, B_{22}) & \cdots & (R_{2n}, G_{2n}, B_{2n}) \\
    \vdots & \vdots & \ddots & \vdots \\
    (R_{m1}, G_{m1}, B_{m1}) & (R_{m2}, G_{m2}, B_{m2}) & \cdots & (R_{mn}, G_{mn}, B_{mn})
    \end{array} 
\right]
\] 
\begin{figure}[H]
    \centering
    \includegraphics[width=9cm,height=5cm]{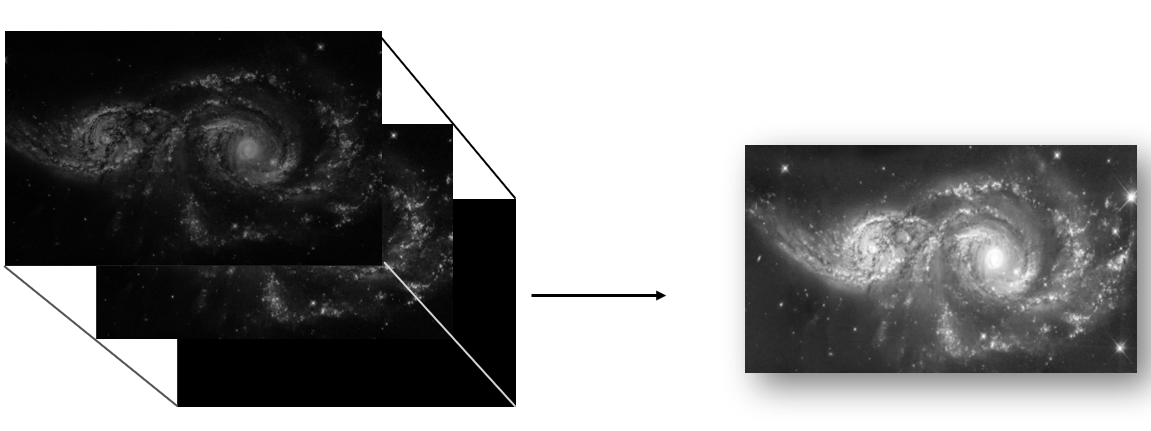}
    \caption{Grayscale Stacked Image}
    \label{fig3}
\end{figure}
So, each element in the stacked matrix is a triplet $(R_{ij}, G_{ij}, B_{ij})$, representing the red, green, and blue intensities for pixel $(i,j)$. After stacking, the 3D matrix $\text{RGB}$ will look like:
\[
\text{RGB}_{ij} = \left( 
\begin{array}{c}
R_{ij} \\
G_{ij} \\
B_{ij}
\end{array}
\right)
\]

\end{justify}

\subsection{Computing the Intensity}

The intensity at each pixel is computed by multiplying the red, green, and blue intensities by their respective constants and summing them up. The weighted sum formula is a linear combination of the red, green, and blue channels, where each channel's intensity is multiplied by a constant factor. This factor acts as a weight, representing the contribution of each color to the overall perceived brightness (intensity). 

\begin{figure}[H] 
\centering
\includegraphics[width=9cm,height=5cm]{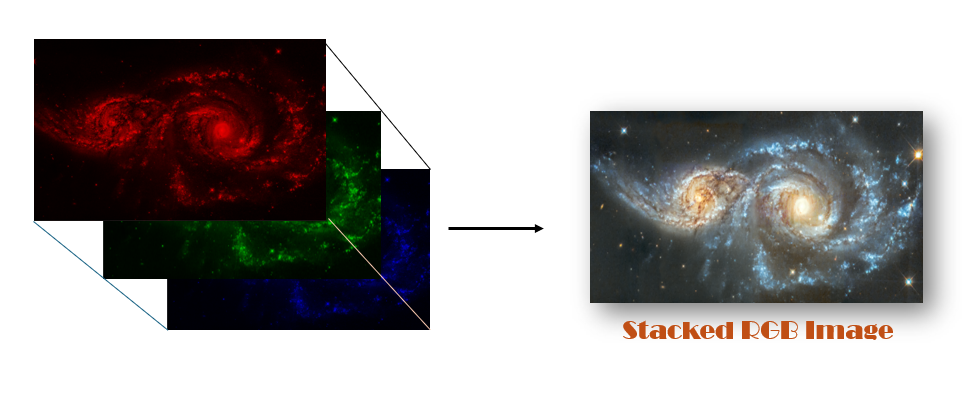}
\caption{RGB Stacked Image}
\label{fig4}
\end{figure}

For an entire image, let’s assume the stacked 3D RGB matrix is denoted as $\text{RGB}$, and each pixel $(i,j)$ has the values $(R_{ij}, G_{ij}, B_{ij})$. We can express the intensity image (Fig~\ref{fig4}) $I$ as:
\[
I_{ij} = \left(
\begin{array}{ccc}
0.299 & 0.587 & 0.114
\end{array} 
\right)
\cdot 
\left(
\begin{array}{c}
R_{ij} \\
G_{ij} \\
B_{ij}
\end{array}
\right)
\]
\newline
Where:
\begin{itemize}
    \item $R_{ij}$, $G_{ij}$, and $B_{ij}$ are the red, green, and blue intensities at pixel $(i,j)$ in the normalized image.
    \item The constants $0.299$, $0.587$, and $0.114$ represent the relative contributions of red, green, and blue to the perceived brightness, based on human visual perception.
\end{itemize}
These weights are chosen based on empirical studies of human vision, specifically derived from the luminance calculations used in video and image compression (\cite{ref12}).

\subsection{Noise Reduction Using Median Filter}

The median filter works by considering a neighborhood of pixels around each pixel \( (i,j) \). For a \( 3 \times 3 \) neighborhood, the median value replaces the center pixel. For the pixel \( R_{ij} \) in the red channel, the filter examines the  neighboring pixels:

\[
\left[
\begin{array}{ccc} 
R_{i-1,j-1} & R_{i-1,j} & R_{i-1,j+1} \\
R_{i,j-1}   & R_{i,j}   & R_{i,j+1}   \\
R_{i+1,j-1} & R_{i+1,j} & R_{i+1,j+1}
\end{array}
\right]
\]

The filter sorts these 9 pixel values and selects the median value. This median replaces the value of \( R_{ij} \)  This process is repeated for every pixel in the red, green, and blue matrices independently.

Let us have an example with a simple 3x3 matrix, \\

\[
\left( 
\begin{array}{ccc}
1 & 2 & 3 \\
4 & 8 & 6 \\
7 & 5 & 9
\end{array} 
\right)
\]
\\

\textbf{Step 1: Handling Pixel \( (1, 1) \)
} \\

This is the top-left corner. Since it's at the edge, we have to pad the surrounding pixels. A \( 3 \times 3 \) window around pixel \( 1 \) (assuming repeated edge pixels): \\

\[
\left( 
\begin{array}{ccc}
1 & 1 & 1 \\
1 & 2 & 2 \\
4 & 4 & 8
\end{array} 
\right)
\]
\\

Sorted values: \([1, 1, 1, 1, 2, 2, 4, 4, 8]\)

Median: \( 2 \)

Replace pixel \( 1 \) with \( 2 \). \\

\textbf{Step 2: Handling Pixel \( (1, 2) \)
}\\

\[
\left( 
\begin{array}{ccc}
1 & 2 & 3 \\
1 & 2 & 3 \\
4 & 8 & 6
\end{array} 
\right)
\]
\\ 

Sorted values: \([1, 1, 2, 2, 3, 3, 4, 6, 8]\)

Median: \( 3 \)

Replace pixel \( 2 \) with \( 3 \). 

\textbf{Step 3: Handling the Center \( (2,2) \)}

A \( 3 \times 3 \) window around pixel \( 8 \): 

\[
\left( 
\begin{array}{ccc}
1 & 2 & 3 \\
4 & 8 & 6 \\
7 & 5 & 9
\end{array} 
\right)
\] 

Sorted values: \([1, 2, 3, 4, 5, 6, 7, 8, 9]\)

Median: \( 5 \)

Replace pixel \( 8 \) with \( 5 \).

If we continue this process applying the Median Filter to each pixel, we will get: 
\\
        \centering
        \textbf{Original Matrix:}
        \[
        \left( 
        \begin{array}{ccc}
        1 & 2 & 3 \\
        4 & 8 & 6 \\
        7 & 5 & 9
        \end{array} 
        \right)
        \]

        \centering
        \textbf{Final Matrix:}
        \[
        \left( 
        \begin{array}{ccc}
        2 & 3 & 3 \\
        4 & 5 & 6 \\
        7 & 7 & 8
        \end{array} 
        \right)
        \]

\begin{table}[H]
\centering
\caption{Comparison of Original and Final Image Values} \label{tab:median_filter_comparison}
\begin{tabular}{|p{2.5cm}|p{2.5cm}|p{2.5cm}|}  
\hline
\textbf{Position} & \textbf{Original Value} & \textbf{Final Value (After Median Filter)} \\
\hline
(1, 1) & 1 & 2 \\
(1, 2) & 2 & 3 \\
(1, 3) & 3 & 3 \\
(2, 1) & 4 & 4 \\
(2, 2) & 8 & 5 \\
(2, 3) & 6 & 6 \\
(3, 1) & 7 & 7 \\
(3, 2) & 5 & 7 \\
(3, 3) & 9 & 8 \\
\hline
\end{tabular}
\end{table}

\subsection{Sharpening the Image using Unsharp Masking}
\begin{justify}

Imagine an image as a 2D surface, where the height of the surface at any point represents the intensity of the pixel at that location.

\begin{itemize}
    \item High-frequency components (edges) are like sharp, abrupt changes in the surface (steep slopes).
    \item Low-frequency components are smooth, gradual changes (flat or gently sloping surfaces).
\end{itemize}
    
\end{justify}

\subsection*{ Blurring with a Gaussian Filter}
\begin{justify}

When we apply a Gaussian filter, our program smooths out the sharp changes, effectively flattening the surface. This removes the high-frequency components (the steep slopes or sharp edges) and keeps only the low-frequency components (smooth transitions). 
For more effective smoothing of images, the Gaussian filter is applied \cite{ref2}. It is the first step of noise remover detection, but not more effective for removing salt and pepper noise. It is based on Gaussian distribution.

The Probability Density Function (P(x)) of Gaussian distribution is represented by Equation (1):

\begin{equation*}
P(x) = \frac{1}{\sqrt{2\pi\sigma^2}} e^{-\frac{(x-\mu)^2}{2\sigma^2}} 
\end{equation*}

Here, \(x\) is a gray level image, \(\mu\) is the mean value, and \(\sigma\) is the standard deviation. The standard deviation (\(\sigma\)) of the Gaussian determines the amount of smoothing.

\begin{itemize}
    \item Visually, this is like flattening out the sharp details in the image while keeping the overall structure.
\end{itemize}
    
\end{justify}
\subsection*{Creating the Mask}
\begin{justify}

The mask is created by subtracting the blurred image from the original image.

\begin{itemize}
    \item Practically, this is like isolating the sharp slopes or edges on the surface. The mask highlights the regions where there were sharp changes (high-frequency components) by calculating the difference between the original and blurred surfaces.
\end{itemize}
    
\end{justify}
\subsection*{Adding the Mask Back}

Finally, this step adds the mask back to the original image. This step amplifies the sharp features (edges) while preserving the smooth areas.

\begin{itemize}
    \item Which is like re-emphasizing the steep slopes in the original surface by increasing their magnitude (strength).
\end{itemize}
Given an image matrix \( I \), unsharp masking can be expressed mathematically as:

\[
I_{\text{sharpened}} = I + \text{strength} \times (I - I_{\text{blurred}})
\]
\\
Where:
\begin{itemize}
    \item \( I \) is the original image.
    \item \( I_{\text{blurred}} \) is the blurred version of the image (low-pass filtered).
    \item \text{Strength} controls the amount of high-frequency detail added back.
\end{itemize}

For each pixel \( I(x,y) \), this operation can be written as:
\begin{align*}
I_{\text{sharpened(x, y)}} 
&= I_{\text(x, y)}\\
&\quad + \text{strength} \times [I_{\text{(x, y)}} - I_{\text{blurred(x, y)}}]
\end{align*}

Thus, the unsharp masking enhances sharpness by increasing the contrast along edges without overly affecting the smooth areas of the image.

\subsection{Intensity Distribution }
\begin{justify}

To analyze the color intensity we also calculated histograms to examine the distribution of the fusion; we considered the distributions of the red, green, blue color distributions. Histogram shows relative density of pixel intensity level and can be used to represent an overall image color density.
Individual pixel intensity was obtained for each channel and histogram of the intensity values was created to show the appearance frequency. To display the distribution we took a histogram  with bins of size 100 and then fitted a Gaussian density distribution on the histogram to represent a smooth function for the intensity distribution. This was done by fitting a Gaussian function to the histogram data points, characterized by the following equation:

\[
f(x) = a e^{-\frac{(x-\mu)^2}{2\sigma^2}} 
\]

where 'f(x)' is the fitted Gaussian function, 'a' is  the amplitude (peak height), $\mu$ is the mean intensity (center of the distribution),
 $\sigma$ is the standard deviation (controls the width of the curve).
    
\end{justify}
\subsection{Framework for Creating Multiple Wavelength Composite Image using Blending }

Creating a composite image from multiple wavelength data, including X-ray, infrared (IR), and optical (RGB) images, then scaling and combining these into a single composite image where each wavelength contributes to a different color channel. Each file (\texttt{.fits}) contains pixel intensity values corresponding to different wavelengths:

\begin{itemize}
    \item \textbf{X-ray}: High-energy radiation image.
    \item \textbf{IR (Infrared)}: Longer wavelength radiation (compared to visible light).
    \item \textbf{RGB (Optical)}: Red, Green, and Blue channels represent visible light. \\
\end{itemize}

Each image can be thought of as a matrix where:

\[
X = \left[ 
\begin{array}{cccc} 
X_{1,1} & X_{1,2} & \cdots & X_{1,n} \\
X_{2,1} & X_{2,2} & \cdots & X_{2,n} \\
\vdots  & \vdots  & \ddots & \vdots \\
X_{m,1} & X_{m,2} & \cdots & X_{m,n}
\end{array}
\right]
\]

Here, \( X_{i,j} \) represents the pixel intensity at position \( (i,j) \) for the X-ray data, and similarly for the IR and RGB channels.

To bring all the images into a uniform intensity scale (range [0, 1]), the pixel intensities are rescaled using the \texttt{exposure.rescale\_intensity} function. For each image (e.g., $X_{\text{data}}$ for X-ray):
\[
X_{\text{norm}} = \frac{X_{\text{data}} - X_{\text{min}}}{X_{\text{max}} - X_{\text{min}}}
\]

This operation normalizes the pixel values, ensuring that the minimum value becomes 0 and the maximum value becomes 1, thus ensuring compatibility between images with different intensity ranges. 

\subsection*{Input Variables:}
\begin{itemize}
    \item $X_{\text{data}}$: The original pixel intensity at some position $(i,j)$.
    \item $X_{\text{min}}$: The minimum intensity value in the original data.
    \item $X_{\text{max}}$: The maximum intensity value in the original data.
\end{itemize}

\subsection*{Subtraction Step:}
\[
X_{\text{data}} - X_{\text{min}}
\]

This shifts the entire range of the data so that the smallest value in the original dataset becomes 0. It shifts all values downward by subtracting $X_{\text{min}}$. \\

When $X_{\text{data}} = X_{\text{min}}$, the numerator becomes 0:
\[
X_{\text{min}} - X_{\text{min}} = 0
\]
So, the normalized value at the original minimum is:
\[
X_{\text{norm}} = 0
\]

This ensures that the minimum value in the data becomes 0. \\

\subsection*{Division Step:} 
\[
\frac{X_{\text{data}} - X_{\text{min}}}{X_{\text{max}} - X_{\text{min}}}
\] \\

By dividing by $X_{\text{max}} - X_{\text{min}}$, the range of the data is scaled so that the original maximum value becomes 1. When $X_{\text{data}} = X_{\text{max}}$, the numerator becomes:
\[
X_{\text{max}} - X_{\text{min}}
\]
So, the normalized value at the original maximum is:
\[
X_{\text{norm}} = \frac{X_{\text{max}} - X_{\text{min}}}{X_{\text{max}} - X_{\text{min}}} = 1
\]

This ensures that the maximum value in the data becomes 1. \\

\subsection*{Intermediate Values:} 

Any value of $X_{\text{data}}$ between $X_{\text{min}}$ and $X_{\text{max}}$ will be scaled proportionally between 0 and 1. \\ 

For example, if $X_{\text{data}}$ is exactly halfway between $X_{\text{min}}$ and $X_{\text{max}}$, say $X_{\text{data}} = \frac{X_{\text{max}} + X_{\text{min}}}{2}$, then the normalized value will be:
\[
X_{\text{norm}} = \frac{\frac{X_{\text{max}} + X_{\text{min}}}{2} - X_{\text{min}}}{X_{\text{max}} - X_{\text{min}}} = \frac{X_{\text{max}} - X_{\text{min}}}{2(X_{\text{max}} - X_{\text{min}})} = 0.5
\]

This means that the data is now scaled so that intermediate values fall between 0 and 1. \\

\subsection*{Intensity Scaling}
\[
X_{\text{scaled}} = X_{\text{norm}} \times \text{Xray\_intensity}
\]

Similar scaling applies for IR and RGB channels.
\[
I_{\text{scaled}} = I_{\text{norm}} \times \text{ir\_intensity}
\]
\[
\text{RGB}_{\text{scaled}} = \text{RGB}_{\text{norm}} \times \text{rgb\_intensity}
\]

\subsection*{Colored Representations for X-ray and IR}

Each wavelength is given a color representation by combining its pixel intensities into specific color channels.

\subsection*{X-ray Representation}
X-ray data is turned into a pink/purple hue by distributing its scaled intensity into the Red, Green, and Blue channels:
\[
X_{\text{colored}} = \left( X_{\text{scaled}} \times 1.0, X_{\text{scaled}} \times 0.3, X_{\text{scaled}} \times 0.6 \right)
\]
This gives higher weight to the red and blue channels, resulting in a pink/purple representation.

\subsection*{IR Representation}
IR data is colored green by setting the Green channel as dominant:
\[
I_{\text{colored}} = \left( I_{\text{scaled}} \times 0.2, I_{\text{scaled}} \times 1.0, I_{\text{scaled}} \times 0.2 \right)
\]

This gives the IR data a strong green hue with subtle red and blue components. \\

\subsection*{Combining the X-ray, IR, and RGB Channels ( Image Blending )}

The composite image is created by adding the scaled X-ray, IR, and RGB data together. Let \( R_{\text{scaled}}, G_{\text{scaled}}, B_{\text{scaled}} \) represent the scaled RGB data. Then, the combined image is given by:
\[
\text{Composite} = \text{RGB}_{\text{scaled}} + IR_{\text{colored}} + X_{\text{colored}}
\]
\begin{figure}[H] 
\centering
\includegraphics[width=9cm,height=7cm]{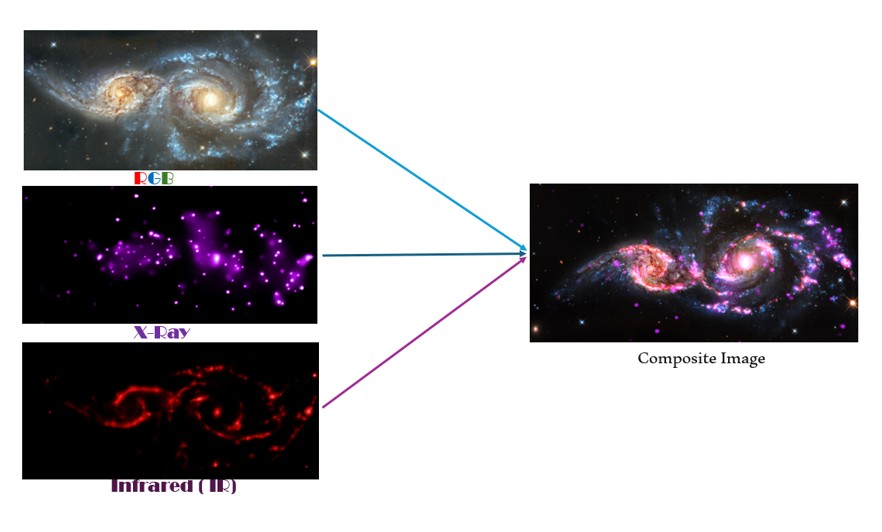}
\caption{Combination of Three Images}
\label{Fig5}
\end{figure}

This operation is performed element wise, adding the pixel values from each component to form a composite pixel (Fig~\ref{Fig5}) for each position \( (i,j) \). \\

\section{Results \& Discussion}

\begin{figure}[H]
    \centering\includegraphics[width=19.3cm,height=7.5cm]{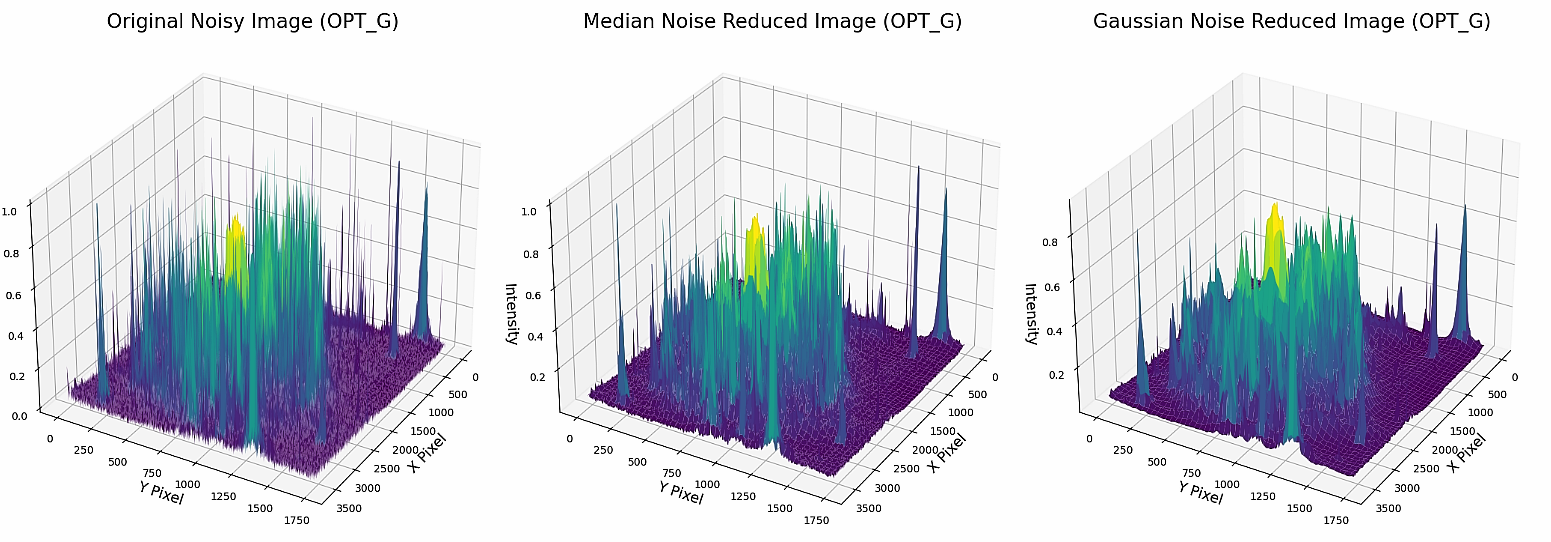}
    \caption{3D plot for Comparing the Original Image with Two Noise Reduced Images for \texttt{OPT\_G.fits}.}
    \label{fig6}

\begin{figure}[H]
    \centering
    \includegraphics[width=19.3cm,height=7.5cm]{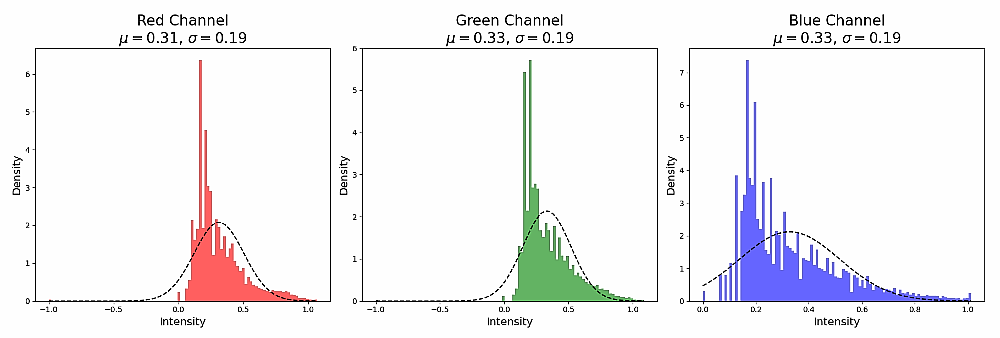}
    \caption{Gaussian Fits and Histograms of Color Intensity Distributions}
    \label{fig7}   
\end{figure}
\end{figure}

\begin{justify}
We successfully created a composite RGB image of an astronomical object using data from three optical filters: blue, green, and red. To obtain normalized images and possible identification of weak signals that were practically invisible earlier, each filter’s data was processed. This processing represents the variation in the level of opacity within the RGB image and enhances small details.\\
\newline
\newline
To further improve image sharpness, a median filter was applied after forming the RGB composite. This noise reduction technique effectively preserved many structural elements in the image foreground while removing random pixel disturbances that could obscure important details. To remove noise, we applied the unsharp masking technique. This method remarkably improved the details. The impact of these processing techniques on image quality is displayed in the 3D plot shown in  (Fig~\ref{fig6}). This comparison expresses improvements in image sharpness and detail following the application of noise reduction and sharpening methods. Using Gaussian fit for the red, green, and blue channels histograms (Fig~\ref{fig7}) represent a uniform distribution of intensity levels across all three channels, indicating a well balanced composite image and effective processing of the individual color channels. After this, we produced a multi-wavelength composite image by combining RGB data with X-ray and infrared (IR) views.
\newline
\newline
Finally, our composite image is merges from multiple wavelengths of data and makes it easier for astronomers to visualize details and inner view of objects based on different parts of the electromagnetic spectrum in a single picture. Composite images are very effective for visualizing the data and extracting various information
within a short time. Therefore, astronomical data analysis is an important step in processing raw
astronomical data, this process provides meaningful interpretations. Here, we used Python for
our astronomical data analysis. Python is an excellent tool for astronomical data analysis, as it is
open source and has a vast collection of implemented packages that can be easily installed on local
machines.
\end{justify}

\section*{Acknowledgements}
\begin{justify}
We would like to thank Research and Project sector of the Copernicus Astronomical Memorial of SUST (CAM-SUST) for arranging the "Astronomy Project for Amateurs" program. Also, we are grateful to Chandra X-ray Observatory for the necessary data as well as the materials for this project.
\end{justify}
\section{ Code and Data Availability}
 All the codes used for the project given here
\href{https://github.com/camsustResearchProject/Astronomy-Project-for-Amateurs/tree/main/11thCommitteeProjects/Project_2}{Astronomy-Project-for-Amateurs}

\end{multicols}
\end{document}